# Advances in Artificial Intelligence for Diabetes Prediction: Insights from a Systematic Literature Review

Pir Bakhsh Khokhar[a,a,*,1], Carmine Gravino[b,a] and Fabio Palomba[c,a]

[a]Department of Informatics, University of Salerno, Via Giovanni Paolo II, 132, Fisciano, 84084, Salerno, Italy

## ARTICLE INFO

## ABSTRACT

Diabetes mellitus, a prevalent metabolic disorder, has significant global health implications. The advent of machine learning (ML) has revolutionized the ability to forecast and manage diabetes early on, offering new avenues to mitigate its impact. This systematic review examines 53 articles on ML applications for predicting diabetes, focusing on datasets, algorithms, training methods, and evaluation metrics. Various datasets, such as the Singapore National Diabetic Retinopathy Screening Program, REPLACE-BG, National Health and Nutrition Examination Survey (NHANES), Pima Indians Diabetes Database (PIDD), are explored, highlighting their unique features and challenges like class imbalance. The review assesses the performance of various ML algorithms such as Convolutional Neural Networks (CNN), Support Vector Machines (SVM), Logistic Regression, and XGBoost for the prediction of diabetes outcomes from multiple data datasets. Techniques such as cross-validation, data augmentation, and feature selection are discussed in terms of their influence on the versatility and robustness of the model. Some of the evaluation techniques involve k-fold cross-validation, external validation, and performance indicators such as accuracy, Area Under Curve, sensitivity, and specificity are presented. The findings highlight the usefulness of ML in addressing the challenges of diabetes prediction, the value of sourcing for different data types, the need to make models explainable, and the need to keep models clinically relevant. The study highlights significant implications for healthcare professionals, policymakers, technology developers, patients, and researchers, advocating for interdisciplinary collaboration and ethical considerations when implementing ML-based diabetes prediction models. By consolidating existing knowledge, this SLR outlines future research directions aimed at improving diagnostic accuracy, patient care, and healthcare efficiency through advanced ML applications. This comprehensive review contributes to the ongoing efforts to utilize AI technology for better diabetes prediction, ultimately aiming to reduce the global burden of this widespread disease.

## 1. Introduction

Diabetes mellitus is a metabolic disorder characterized by elevated blood glucose levels due to either insufficient insulin production by the pancreas or improper insulin utilization by the body. Insulin, a crucial hormone secreted by the pancreas, facilitates the movement of glucose from the blood into the cells, where it is converted into energy. It is also essential for the metabolism of proteins and lipids. When the body does not produce enough insulin or the cells do not respond to it properly, glucose accumulates in the blood, leading to diabetes.

This condition can result in severe complications such as heart disease, kidney failure, and nerve damage. According to the International Diabetes Federation, the number of people with diabetes worldwide is projected to reach 700 million by 2045 [1]. This alarming prediction underscores the urgent need for innovative methods and treatments for diabetes.

Traditional diabetes treatments focus on monitoring blood glucose and HbA1c levels, which are reactive approaches that detect the disease at an advanced stage. Thus, the need to develop better models for early prediction to improve the quality of life of a patient cannot be overemphasized. Studies have also shown the revolution that has been brought in the healthcare industry by AI, especially in ML and DL [24]. These technologies are particularly good at capturing large amounts of data, as well as recognizing patterns and even making predictions that were inconceivable before the use of such statistical tools. Since the interest in applying the ML for predicting diabetes has been on the rise, research in this field has received a boost. The accuracy of developed ML models to predict diabetes relies greatly on the ML model and type and amount of data used, such as Electronic Helath Recors (EHR), laboratory data, age, gender and other aspects of lifestyle [31]. The integration of Continuous Glucose Monitoring (CGM) data with EHRs has been more useful, especially in predicting health outcomes, than the use of CGM data alone [51]. Also the integration of genetic information and biomarkers brings more information about the probability of the person to develop diabetes [25].

Training of the ML models for the prediction of diabetes includes different procedures and optimizations. Logistic regression, SVM and random forest are some of the widely used algorithms in supervised learning because of their interpretability and stability [34]. Other deep learning models, including CNNs and RNNs, have also been used in the analysis

---

[*]Corresponding author
[**]Principal Corresponding author

p.khokhar@studenti.unisa.it (P.B. Khokhar); gravino@unisa.it (C. Gravino); fpalomba@unisa.it (F. Palomba)
https://rubrica.unisa.it/persone?matricola=066182 (P.B. Khokhar); https://docenti.unisa.it/004724/home (C. Gravino); https://docenti.unisa.it/027888/home (F. Palomba)
ORCID(s):

[1]International Diabetes Federation. Accessed July 7, 2024. https://idf.org/about-diabetes/diabetes-facts-figures/

of the data to establish complex patterns [2]. Practices such as transfer learning and ensemble approaches have become more popular in the effort to improve the generalization and predictive capabilities [81]. The effectiveness of ML models in diagnosing diabetes especially the accuracy of the diagnosis must also be determined for the models to be practical. Some of these assessment metrics include: accuracy, precision, recall, F1 measure, and the AUC-ROC [62]. Sensitivity, specificity and MCC are also employed to measure the accuracy of a model in the identification of true positive and /or true negative results [21].

Systematic literature reviews (SLRs) are useful in offering an integrated analysis of the literature for a given subject area in the field of AI-based diabetes prediction. An SLR allows a researcher to determine trends, gaps and patterns in the use of AI for diabetes prediction, so that techniques and result can be compared to enhance the creation of better ML models [15].

Through this SLR, specific details are highlighted on the current developments in ML-based diabetes prediction with respect to the datasets, training, and evaluations. These dimensions were scrutinized by the authors to determine the state of the art of research in the field, trends in the field, and possibly areas of research that might have not been explored before. It is hoped that the results can help us to find the future work and improve the precision and applicability of ML models to diabetes prediction. The review addressed three main questions:

- First, it discussed the basic data and their properties utilized in the models for the prediction of diabetes and effect of these properties on the predictive models.

- Second, it described conventional training approaches and different ways to improve the accuracy of model and its ability to generalize.

- Lastly, it examined the evaluation criteria for ML models, especially on the commonly used metrics to measure the performance of the model.

Overall, this review provides valuable insights into the current use of ML in predicting diabetes, highlighting technical aspects of model development and practical implications for healthcare. By summarizing current research and identifying key trends and gaps, the review contributes to ongoing efforts to leverage AI technology for improved diabetes care, ultimately aiming to reduce the global impact of this widespread disease and improve patient outcomes.

**Structure of the paper:** Section 2 provides an in-depth look at the anatomy of diabetes, reviewing previous research and existing systematic literature reviews to set the context for the current study. **Section 3** details the stepwise approach of research questions guiding the review, the systematic review methodology, inclusion and exclusion criteria, database search strategies, quality assessment, and data extraction steps. **Section 4** presents the findings from the reviewed studies, analyzing datasets used, machine learning algorithms, training strategies, and evaluation metrics. Further, it interprets the results obtained from the SLR. **Section 5** discusses the limitations, potential biases, and gaps in the literature also provides implications for the researchers and stakeholders. **Section 6** discusses the threats to validity of the work conducted in this study and finally **Section 7** concludes the key findings, reaffirming the potential of machine learning in diabetes prediction and concluding with thoughts on the future of ML in healthcare and its role in improving diabetes prediction.

## 2. Background and Related Work

The main objective of this section is to equip us with background knowledge of the problem so that we can proceed with our research. This section is devoted to the data methods, prediction models, and metrics used for diabetes prediction and to systematic literature reviews conducted in the past regarding diabetes prediction.

### 2.1. Anatomy of Diabetes Mellitus

Diabetes mellitus (DM) is a metabolic disorder in which the body is unable to regulate levels of sugar or glucose in the bloodstream either due to inadequate insulin secretion by the pancreas (type 1) or due to insulin resistance (type 2). There are two main types of diabetes: Type 1 Diabetes Mellitus (T1DM) and Type 2 Diabetes Mellitus (T2DM), with Type 2 diabetes comprising nearly 90% of all diabetes cases and with a global prevalence of 537 million [17]. Diabetes is a community health problem that has shown an alarming increase in the last 20 years in many parts of the world. DM is a multiorgan disease with numerous diabetic microvascular complications involving the retina, heart, brain, kidney, and nerves.

It can be clearly stated that the role of medical personnel in preventing, treating, and managing diabetes mellitus and its complications is well established [5]. Exercise prescription and education for rehabilitation management are effective for participation and maintaining physical well-being, improving situation of patient and health-related quality of life [9]. Diabetes itself is not a high-mortality cause, but it is a significant risk for other causes of death and has a high disability burden. Diabetes is also a significant risk factor for cardiovascular disease, kidney disease, and blindness [12]. DM is categorized into three types according to their etiology and clinical manifestation: type 1 diabetes, type 2 diabetes, and gestational diabetes [29].

Diabetes Mellitus primarily involves the islets of Langerhans in the pancreas, from which glucose is secreted from the alpha cells and insulin from the beta cells. Glucagon increases blood glucose levels, and insulin reduces them. T1DM (Insulin-Dependent) is a chronic metabolic disorder that causes 5% to 10% of diabetes mellitus [26]. It is characterized by the autoimmune destruction of insulin-producing beta cells in the islets of the pancreas, and the loss of function of the beta cells leads to absolute insulin deficiency. T1DM is most commonly seen in children and adolescents but can affect anyone at any age.

T2DM (Non-insulin dependent) comprises 90% of all diabetes [29]. The reduction in the effect of insulin in T2DM is called insulin resistance. Under normal conditions, insulin is ineffective and, therefore, is initially countered by an increase in insulin production to maintain glucose homeostasis but later decreases to cause T2DM. T2DM is common in adults aged 45 years or older [1]. It is now more prevalent in children, adolescents, and younger adults as a result of rising obesity, physical inactivity, and energy-dense diets.

Gestational Diabetes Mellitus (GDM) can occur at any stage of pregnancy. Typically, it happens to pregnant women in the second and third trimesters. The American Diabetes Association (ADA) estimates that GDM occurs in 7% of pregnancies. GDM and its offspring are at elevated risk of developing type 2 diabetes mellitus in the future [76].

## 2.2. Related Work

In previous years, as evidence shows, several systematic literature reviews emphasize the diagnosis of predicting type 2 diabetes and studies concerning those predictions. Many of the articles from those journals and conferences are centered around Machine Learning and Deep Learning techniques, which are among the most relevant topics today. They aim to investigate similar data sets and conclude through the data sets analysis that the amount of data used in those studies is unstable.

The research conducted by Bidwai, P. et al. [11] has suggested a new review that aimed to eliminate the gaps left by current reviews and helped other researchers in selecting the current results from the studies that they can use in predicting ML-based risk of Diabetic Retinopathy progression and related diseases by synthesizing the current results from these studies and put in place the research challenges, limitations, and gaps for the selection of efficient machine learning techniques in the establishment of my model of prediction. Furthermore, they pointed out the six AI-related technical discussions and the approaches as these two crucial points for the adopted strategy. As for the SLR, data collection was used to obtain suitable studies. They searched IEEE Xplore, PubMed, Springer Link, Google Scholar, and Science Direct electronic databases for literature reviews published between January 2017 and 30th April 2023. Thirteen (13) studies appearing from the broad discussion were subsequently shortlisted based on their relevance to the reviewing questions and the filters applied. While the literature review exposed some significant research gaps to be considered in future research that will improve the performance of Diabetic Retinopathy (DR) progression risk prediction models, issues like the comparability and inclusion of the diverse DR populations were inattentive.

They also discussed different approaches to the problem of diabetes prediction in general, and about the problem of selecting and integrating necessary research articles for ML-based diabetic prediction models. They talked about how the medical data is nonlinear, non-normal and correlation structured and about how beneficial machine learning is in healthcare especially in the medical imaging. While their review was not comprehensive in some of the areas of interest especially in early diagnosis and risk stratification, it provided the researchers with a source of reference. However, the current systematic literature review (SLR) follows the PRISMA guidelines much more closely to ensure more exhaustive and objective approach to analysis and provide the discussion of the practical recommendations for further research that would consider the intricacies of medical data for diabetes prediction.

It may preclude older basic studies because the ML-based risk prediction of DR progression [74] is limited to papers published between January 2017 and April 2023. Using only 13 research and a few databases may not identify all the relevant materials, which can lead to selection bias. The authors did not extensively discuss the comparability and inclusion of the different DR populations, which would influence the generalizability of the findings. Our SLR alleviates these limitations by focusing on a more extended period (2014-2023), covering more first-hand papers (53), and incorporating more criteria such as algorithms, datasets, and validation methods. This methodological approach increases the likelihood of identifying relevant and inclusive studies. It thus provides a more comprehensive synthesis of the literature as a foundation for future research on blood glucose prediction and DR progression.

The systematic literature review performed by Wadghiri, M.Z et al. [76] aimed to review state of the art in predicting blood glucose using ensemble methods regarding eight criteria: types of algorithms, year of publication, journal, database, types of ensembles, learners, combination methods, performance measures, validation methods, overall performance, and accuracy. This systematic literature review has been performed to compare primary studies between digital libraries from 2000-2020. Among the 32 primary papers they have reviewed, eight review questions were chosen for this study. The results indicated an increase in the use of ensembles in recent years; overall, they were better than the rest of the single models. However, the process of formation of the groups and the performance criteria are not entirely flawless. Here, some suggestions have been provided about the design of compelling ensembles for blood glucose level prediction.

Digital libraries may have missed crucial studies. The exclusion of some research and a short number of evaluated primary papers may affect the selection process comprehensiveness and biases. The study approach of measuring ensemble formation and performance has limitations. This discovery is particular to blood glucose prediction and may not apply to other contexts. Datasets and validation methods also affect dependability. Finally, the pace of technological progress may render certain conclusions outmoded and less relevant. This article addresses these concerns well, focusing on significant database research and publication years. We use various methods besides ensemble learning.

The review by Eijoseno, M.R et al. [77] was designed to present diabetes in general, its prevalence, complications, and opportunities for artificial intelligence in early diagnosis and classification of diabetic retinopathy. The research also focused on ML-based methods like machine learning and deep learning. The new research areas that include transfer learning using generative adversarial networks, domain adaptation, multitask learning, and explainable artificial intelligence in diabetic retinopathy were also considered. A list of the methods already in use, the screening systems, the performance measurement, the biomarkers in diabetic retinopathy, the potential issues, and the challenges in ophthalmology was presented. The future scope was elaborated on in the conclusion. The review may lack systemic rigor because it focuses on diabetes and ML methods without using Preferred Reporting Items for Systematic reviews and Meta-Analyses (PRISMA). Only some powerful ML algorithms may be included, while others are omitted. Also, the assessment may not give immediate practical suggestions while planning for future work. Our SLR is more rigorous and systematic because it conforms to the PRISMA framework. It also includes more criteria and approaches and provides a more comprehensive analysis and application recommendations for future research on AI-based prediction.

The comprehensive review by Saxena, R. et al. [67] presents the current literature on machine learning in diagnosing DM. The research dealt with the use of machine learning models and datasets for the diagnosis of diabetes. The results showed how Random Forest can be used successfully and how it is prevalent in this area of research. A prompt diagnosis of diabetes is essential because it helps control the disease and avoid complications. Nevertheless, the fact that people have no access to care and that there are cases that go undiagnosed are also challenges. The analysis presented problem areas such as data quality, sensitivity-specificity trade-offs, incorrect readings, and missing data. The authors have further said that future research must be expanded by enlarging the training data set, including additional parameters, and addressing the outlier handling methods to overcome these challenges. Moreover, feature selection methods and the issue of which is more critical, sensitivity or specificity, should be considered. Although this process has problems, machine learning can make diabetes detection easier and improve medical care. Therefore, the present research gives future researchers a chance to know more about implementing ML algorithms for diabetes diagnosis.

This review has listed the following drawbacks in ML for DM diagnosis: Random Forest is practical and widely used in this field, but the assessment identifies data quality issues, the sensitivity-specificity curve, false readings, and incomplete records. The paper also suggests more significant training datasets, parameters, and better outlier handling. It also implies improved feature selection and a better understanding of the relationship between sensitivity and specificity. However, our SLR aims to overcome these constraints by being more inclusive. It enforces the PRISMA framework for systematic data gathering and analysis, encompasses several machine learning techniques and considerations, and provides actionable research recommendations. This comprehensive strategy improves ML for diabetes diagnosis and solves emphasized issues.

The systematic review of the literature on data-driven algorithms and models was performed by Felizardo, V. et al. [33] using accurate diabetic data to predict hypoglycemia. The review process was an intense one that spanned over five electronic databases ScienceDirect, IEEE Xplore, ACM Digital Library, SCOPUS, and PubMed covering publications from Jan 2014 to Jun 2020. This search yielded 63 studies included in the analysis due to their relevancy. The review showed that data models developed for predicting blood glucose and hypoglycemia might have to balance applicability and performance. This formed the integration of other data sources or using different modeling approaches. The study outcomes proved the current tendencies and prompted further research on hypoglycemia prediction. This systematic analysis of data-driven hypoglycemia prediction comprises 63 articles from five databases from 2014 to 2020. Though comprehensive, its brief timespan may omit recent developments. It may not pay much attention to data variety and the combination of its distinct techniques to focus more on the applicability and performance of the model.

El Idrissi et al. [40] attempted mapping and reviewing the existing literature that explored the use of data mining (DM) predictive techniques in diabetes self-management (DSM). In their review, they preferred 38 papers that were published between the years 2000 and April 2017 to categorize and review the literature on the application of DM techniques for DSM tasks including blood glucose level prediction, hypoglycaemia detection and insulin dose estimation. The review established that artificial neural networks were the most popular kind of predictive technique and the second was the auto-regressive type of models and the third was the support vector machines. Interestingly, the majority of investigations concerned T1DM the most frequent clinical issue was blood glucose prediction which was the target of more than 57% of the selected investigations.

The authors also highlighted some of the issues, including the lack of model generalization as a result of patient-specific data, high complexity involved in regulating blood glucose levels, and variations in metrics used in the assessment of results across the studies. Nevertheless, the review pointed out that DM techniques such as ANNs and autoregressive models could hold significant future capacity for enhancing DSM prediction accuracy and decision-making. Nevertheless, the study called for more research in the use of hybrid models, and the extension of these techniques in the T2DM and gestational diabetes; also, the need for a more standardized experimental design in future research in this field was highlighted.

Our SLR employs PRISMA for further rigor and coverage. It includes research conducted from the foundational to the contemporary period and covers 2014–2023. Compared with evaluating model performance, our evaluation involves algorithms, datasets, validation, and challenges for blood glucose prediction, which provides a more comprehensive and applicable perspective for this research.

These papers aim to provide an in-depth discussion of diabetes mellitus and include discussions on the various types of diabetes, the number of people affected by diabetes, and the different health complications associated with diabetes. They emphasize the importance of systematic reviews and ML-based strategies when studying the use of ML and deep learning (DL) technologies for the effective prediction and management of diabetes through the analysis of the application of these technologies.

However, several limitations are still present, including data quality issues, system interoperability challenges, and disease classification. These limits underline the fact that continuous innovation in this discipline is necessary. The research emphasizes the importance of developing current predictive models, exploring novel approaches of artificial intelligence, and utilizing various data sources to enhance the efficiency and accuracy of diabetes prediction tools, respectively. To overcome these constraints, there is a need to improve the quality of data, establish better approaches for system integration, and improve classification algorithms to create more effective and applicable artificial intelligence models in diabetes prediction.

## 3. Research Approach

The predominant goal of this SLR is to achieve critical systematic integration and summary of the latest published scientific literature on the application of machine learning in predicting and managing diabetes. This review will outline, analyze, and summarize emerging trends, known gaps, and key takeaways of the technology landscape that is dynamic and fast-moving. This review aims to examine the predictive models; additionally, it will outline the used approaches, both the strengths and the weaknesses, used datasets, training and validation strategies, categorize the effectiveness of the current hypothesis, give the critique, and consider the areas to advance further research. To achieve this, the review process must be thoroughly arranged according to the PRISMA framework [59]. With its solemnity and complexity of procedure, PRISMA is considered a reference to conducting systemic reviews, supporting hearings, and ensuring clarity in the appraisal of scientific literature. It provides a systematic technique that is evaluative regarding literature selection, literature assessments, and literature syntheses. This makes it a proper analysis tool that condenses vast research findings into coherent conclusions.

### 3.1. Research Objectives and Research Questions

The research questions designed for the systematic literature review aim to answer how machine learning and artificial intelligence are used for diabetes prediction and establish the framework for the current state of the art in the field. These goals are not only to sum up and analyze the previous studies but also to identify the areas of gaps where more technological innovations and new methods can be used. The main objective is a systematic review of all the possible areas of the application of machine learning and artificial intelligence technologies in diabetes prediction to create a framework for understanding the limitations of what is technically feasible and clinically applicable. The objectives are as follows:

**Objective 1:** *To identify and synthesize the findings on datasets with their characteristics utilized in diabetes prediction.*

**Objective 2:** *To examine the configurations and the range of ML techniques used in diabetes prediction.*

**Objective 3:** *To analyze evaluation setups and performance metrics used in ML models to predict diabetes.*

**Objective 4:** *To identify the limitations of current research in diabetes prediction.*

These objectives are methodically developed to ensure that the systematic review is comprehensive, based on hard evidence, and relevant to the current progress of health technologies. The goal is to effectively connect the theoretical research to the practical application of the observations gained in this review to enable the people on the front lines of diabetes prediction to act accordingly.

Through the systematic fulfilment of these target areas, the review is designed to be a pivotal tool that helps various stakeholders. It intends to inform and enrich knowledge of researchers by giving a comprehensive summary of the existing methodologies and the results they have generated, shedding light on the successful techniques and the areas that require further exploration. For care providers, this review is meant to transform data-driven insights into clear and actionable steps that can be used to improve the health of patients. Furthermore, the policymakers have the basis of evidence synthesized to support the decision-making process concerning the policies that guide healthcare, financial allocations, and strategic planning in diabetes care.

**RQ$_1$.** *What datasets, including their characteristics, have been utilized in research studies focused on diabetes prediction?*

This research question is to discover and explain the datasets that have been used in the studies that have been focusing on diabetes prediction. Through this process, we can estimate the size of the data, the diversity, and the representativeness of the population, which are crucial for developing a robust and applicable model to different populations. Additionally, analyzing these datasets will also show any deficits in data utilization that could be corrected in future studies, thereby contributing to an improvement in the accuracy and the generalizability of diabetes diagnostic tools. Through this study, we will set data standards for diabetes prediction research and thus provide a basis for other studies to build upon the foundations of such data.

By addressing **RQ$_1$**, we will be able to meet the **Objective 1**. Which is to explore the datasets and their characteristics used for predicting diabetes. Understanding these elements helps us to assess the current data standards and identify potential gaps in data utilization.

> **RQ$_2$**. *What are the configurations of ML approaches used in diabetes prediction, including the independent variables, classification types, ML algorithms, and training strategies?*

This extensive research topic aims to investigate the peculiarities of the application of artificial intelligence tools in diabetes prediction. It seeks to understand the various components that contribute to the development and optimization of ML models in this context. This includes identifying the independent variables considered influential in predicting diabetes, such as patient demographics, health metrics, and genetic information. Additionally, it explores the classification types used to differentiate between outcomes, such as distinguishing between types of diabetes or predicting disease progression stages. The question also investigates the range of AI algorithms—from traditional machine learning to advanced deep learning techniques—harnessed to analyze and interpret complex datasets effectively. Lastly, it examines the training strategies implemented to enhance model performance, including methods for training data selection, model validation, and techniques to prevent overfitting. Understanding these aspects provides a clear picture of the current state of AI applications in diabetes prediction and identifies areas for potential improvement and innovation. By examining the configurations of ML approaches, use of independent variables and finding classification types, we will be able to meet the **Objective 2**.

> **RQ$_3$**. *What are the various evaluation setups employed in the context of diabetes prediction, explicitly focusing on the types of validation methods used and the metrics applied to assess the effectiveness of these models?*

This research question seeks to explore and characterize the evaluation frameworks used in the context of diabetes prediction, emphasizing the methodologies applied for validating predictive models and the metrics used to measure their effectiveness. It aims to understand the diversity and robustness of validation techniques, such as cross-validation, bootstrapping, or external validation, which ensure the reliability and generalizability of ML-driven prediction models. In addition, the inquiry analyzes the specific performance metrics utilized in evaluating these models. These metrics include accuracy, sensitivity, specificity, AUC-ROC, and several others. By analyzing these aspects, the research can identify best practices and potential areas for enhancement in the assessment of ML models, contributing to improved outcomes in diabetes prediction. **RQ$_3$** plays a crucial role in fulfilling **Objective 3** by providing detailed insights into the setup of ML models. This includes independent variables, classification types, ML algorithms, and training strategies, offering a comprehensive view of how ML tools are tailored to enhance predictive accuracy in diabetes care.

Therefore, this research looks at datasets, ML methods, and training schemes to understand the shortcomings of existing diabetes prediction research, where shortcomings in data quality and characteristics will be highlighted. Others include overfitting, computational burden, and interpretability of the model, among others. The investigation delineates blind spots of the current research and outlines further study directions by fulfilling the **Objective-4**, which will enhance the credibility and generalizability of the diabetes prediction models.

### 3.2. Search Databases and Search Queries

When conducting a systematic literature review, particularly in fields involving advanced technologies such as ML in healthcare, selecting suitable databases to search is crucial. In the context of systematic literature reviews, a research query definition specifies the exact terms, scope, and parameters for a search strategy used to gather relevant literature on a given topic. This definition is crucial as it directly influences the quality, relevance, and comprehensiveness of the literature collected. Defining a research query helps ensure the review is systematic, reproducible, and closely aligned with the research objectives. Keeping in mind that definition, we set the following strategy:

- Specific words and phrases were used in the database search. Those were usually derived from the main topics of the research questions and are critical in retrieving relevant literature. Keywords are carefully selected to capture the various aspects of the investigated topic.

- For all those keywords, we searched for synonyms, alternative spellings, and other names for the disease.

- We incorporated Boolean operators (AND, OR) to formulate search queries.

Here is a description of critical databases with their search queries that were used for such research, highlighting their specific relevance and benefits:

**PubMed** is the premier database for anyone researching medical and healthcare topics. Managed by the National Institutes of Health, PubMed provides access to more than thirty million citations for biomedical literature from MEDLINE, life science journals, and online books. It is especially useful for finding peer-reviewed articles on medical studies, clinical trials, and epidemiology.

**Search Query for PubMed**

(((((data*) OR (variable*)) AND ((diabetes*) OR (diabetes insipidus) OR (diabetes mellitus) OR (polygenic disease) OR (polygenic disorder)) AND ((AI) OR (artificial intelligence) OR (ML) OR (machine learning) OR (deep learning)) AND ((predict*) OR (detect*) OR (identify*) OR (discover*) OR (find*) OR (recogniz*) OR (determin*) OR (anticipat*) OR (project*) OR (estimat*)) AND ((train*) OR (validat*) OR (metric*) OR (evaluat*))))

The PubMed search query focuses on artificial intelligence and machine learning in diabetes research, specifically focusing on data forms and variables. It includes terms on deep learning, predicting, detecting, identifying, and estimating diabetes-related aspects. The query also includes terms on methodologies used, ensuring comprehensive discussions on effectiveness of ML models in diabetes prediction and management.

A critical resource for technology-focused research, **IEEE Xplore** provides access to the content from the Institute of Electrical and Electronics Engineers (IEEE) and the Institution of Engineering and Technology (IET). It includes over four million documents, including articles, conference papers, and standards, essential for research involving technological applications in healthcare like AI and ML algorithms, software, and system implementations.

**Search Query for IEEE Xplore**

((((("data" OR "dataset" OR "variable*") AND ("diabetes" OR "diabetes insipidus" OR "diabetes mellitus" OR "polygenic disease" OR "polygenic disorder") AND ("artificial intelligence" OR "machine learning" OR "deep learning" OR "ML" OR "DL") AND ("predict*" OR "detect*" OR "identif*" OR "discover*" OR "find*" OR "recogni*" OR "anticipat*") AND ("training" OR "validating" OR "validation" OR "matric*" OR "evaluate" OR "evaluating" OR "evaluation" OR "examine" OR "examining" OR "examination"))))

The search query for IEEE Xplore includes various terms related to diabetes from abstracts of the papers which are AI, ML, machine learning, and deep learning, with the aim of predicting, detecting, discovering, finding, recognizing, determining, anticipating, projecting, evaluating, and training.

The **ScienceDirect** is owned by Elsevier and offers various scientific and technical research covering physical sciences, engineering, life sciences, health sciences, social sciences, and humanities. This database is precious for comprehensive searches in interdisciplinary fields that combine technology and healthcare, providing access to a vast library of scientific articles, book chapters, and other resources.

By effectively using these databases, we can access the most relevant and comprehensive information for their systematic reviews in ML applications in diabetes prediction. Each database offers unique tools and collections that can significantly enhance the depth and breadth of a literature review.

**Search Query for Science Direct**

((data*) OR (variable*)) AND ((diabetes*) OR (diabetes insipidus) OR (diabetes mellitus) OR (polygenic disease) OR (polygenic disorder)) AND ((AI) OR (artificial intelligence) OR (ML) OR (machine learning) OR (deep learning)) AND ((predict*) OR (detect*) OR (identif*) OR (discover*) OR (find*) OR (recogniz*) OR (detremin*) OR (anticipat*) OR (project*) OR (estimat*)) AND ((train*) OR (validat*) OR (metric*) OR (evaluat*)))))))

The Science Direct search query focuses on machine learning (ML) in diabetes research, specifically predictive and diagnostic models. It includes keywords related to data handling and variables, ensuring relevance to diabetes and its genetic interactions. The query highlights innovative methods, functional objectives, and methodological aspects, providing insights into current trends, challenges, and advancements in the field.

Due to the restriction of using only 8 Boolean operators in the Science Direct database, we split the Search Query for Science Direct into five sub-queries to search all relevant articles. Later, we filtered out the unique articles and removed duplicates.

**Search Query 1 for Science Direct**

((diabet) OR (diabetes insipidus) OR (diabetes mellitus) OR (polygenic disease) OR (polygenic disorder)) AND ((data) OR (variable))

**Search Query 2 for Science Direct**

((diabet) OR (diabetes insipidus) OR (diabetes mellitus) OR (polygenic disease) OR (polygenic disorder)) AND ((AI) OR (artificial intelligence) OR (ML) OR (machine learning) OR (deep learning))

**Search Query 3 for Science Direct**

((diabet) OR (diabetes insipidus) OR (diabetes mellitus) OR (polygenic disease) OR (polygenic disorder)) AND ((predict) OR (detect) OR (identify) OR (discover) OR (find))

**Search Query 4 for Science Direct**

((diabet) OR (diabetes insipidus) OR (diabetes mellitus) OR (polygenic disease) OR (polygenic disorder)) AND ((recognize) OR (determine) OR (anticipate) OR (project) OR (estimate))

**Search Query 5 for Science Direct**

((diabet) OR (diabetes insipidus) OR (diabetes mellitus) OR (polygenic disease) OR (polygenic disorder)) AND ((train) OR (validate) OR (metric) OR (evaluate))

By effectively using these databases and their search queries, a comprehensive search strategy was crafted to retrieve relevant literature, and we were able to access the most relevant and comprehensive information for systematic reviews of ML applications in diabetes prediction from 2014 to 2023.

### 3.3. Exclusion and Inclusion Criteria

Exclusion and inclusion criteria can facilitate the selection of resources that address the research questions in a systematic literature review. Within the framework of our investigation, we determined and implemented the "Inclusion/Exclusion" criteria that should be followed. For the **Exclusion Criteria** during our research, we eliminated the resources that were able to satisfy the following constraints:

- Articles written in languages other than English.
- Short papers are defined as papers that consist of fewer than seven pages.
- Workshop papers
- Papers that are duplicated.
- The full text of the papers that were not available for reading.
- In subsequent years, conference papers were published in journals.

For the **Inclusion Criteria**, all the articles that applied machine learning methods to predict diabetes were included in our study.

### 3.4. Execution of Search Queries

Once we had the general framework for the SLR in hand, we designed a thorough search strategy to cover all the databases and widen the scope of our search.

**A.** The search yielded many relevant articles from three major databases. Consequently, the data for research were collected from 321 articles from IEEE Xplore, 728 papers from PubMed, and 807 articles from Science Direct. These diversified sources were useful in building up a good pool for the review which will be useful in the development of the database. The first search brought up a total of 1,856 articles in all the databases searched for. On the records that were collected, the process of de-duplication was also done to avoid having the same record entered twice. Finally, after excluding duplicate articles 336, there were 1520 articles that went through the screening process.

**B.** Each manuscript was then proceeded to the exclusion criteria in a step wise manner. In this phase, all records identified by the search process amounted to 1,520, and all the records were screened by title and abstract to decide their suitability for the study. Out of this process, 1468 records were screened out because they were not relevant to the research questions or did not meet the inclusion criteria. Out of them, 53 were potentially relevant and hence were retrieved for full-text review based on the title and abstract.

**C.** The first author of the study systematically went through the 53 manuscripts and strictly obeyed the rules of the inclusion criteria. Thus, 37 studies were included in the analysis after the full-text review of the articles and according to the data quality, the relevance of the studies, and the purpose of this study. Out of the total 37 studies, a total of 16 studies were removed at this step; 11 studies were not related to the research questions and for 5 studies, the required information was missing. The other 37 studies were considered to be of high quality and more related to the systematic review.

**D.** To make the process of identifying relevant articles even more rigorous, the snowballing technique was used. Regarding the citation searching method, it means the use of references or citation from the previous studies that have been included in the current study and help in identifying more related studies that may have been retrieved from the database search. There are two types of snowballing: forward snowballing that entails identifying papers that have cited the included papers and backward snowballing that involves identifying papers cited in the included papers. To ensure the systematic review of the reference lists of the 37 studies, backward snowballing was applied. By using this snowballing method, other 16 related studies were also found, and all of them were included in the final review. These works greatly expanded the range of the literature and greatly reduced the likelihood of missing any pertinent research.

**E.** The last process was the process of incorporating studies. By the end of the study, a total of 53 papers were analyzed in the systematic review after the eligibility step and the snowballing technique were done; this was after getting 37 papers from the eligibility step and an additional 16 papers from the snowballing technique. This approach ensured a consistent and systematical method of selecting the studies as shown in Figure 1, which makes a solid ground for answering the research questions and yielding useful knowledge.

**F.** We progressed to the data extraction stage, which is crucial for answering our study questions by identifying the exact datasets with characteristics, training strategies, evaluation approaches, metrics and ML algorithms used in the studies. The data collection process was easy, enabling the primary author to handle the extraction independently. Yet, assessing the possible constraints of the investigations was more difficult. This analysis required a thorough and concentrated discussion, collaboratively carried out by all authors of our work. They carefully analyzed sections of the publications that addressed potential constraints and threats to validity. They examined the features and qualities of each ML technique employed to pinpoint any further constraints. All authors are experienced in artificial intelligence and machine learning, with years of expertise and engagement in teaching academic courses. This significantly improved the thoroughness and depth of their analysis in this phase of our systematic literature evaluation.

Table 1
Summary of Extracted Articles from Various Databases

| Database | Period | Document Type | Publication Stage | Language | Media Format | Subject of Interest | No. of Papers |
| --- | --- | --- | --- | --- | --- | --- | --- |
| IEEE Xplore | 2014-23 | Conference Proceedings and Journals | Final and Published | ENG | PDF | Computer Science/Engineering | 321 |
| PubMed | 2014-23 | Conference Proceedings and Journals | Final and Published | ENG | PDF | Computer Science/Engineering | 728 |
| ScienceDirect | 2014-23 | Conference Proceedings and Journals | Final and Published | ENG | PDF | Computer Science/Engineering | 807 |
| Total | | | | | | | 1856 |

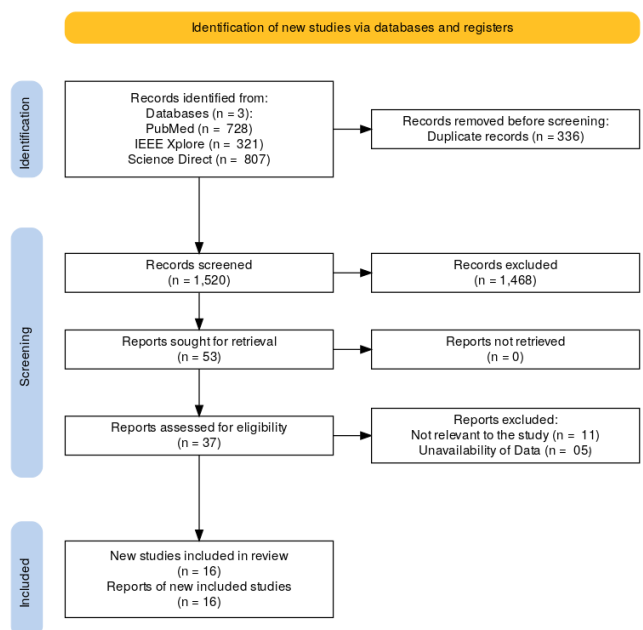

Figure 1: PRISMA flowchart of study identification, screening, and inclusion process.

The information we obtained from the selected publications answered our research questions. This section offers a concise review of the most important findings that emerged from our investigation.

### 3.5. Quality assessment

Before moving further with the process of extracting the material necessary to answer our research questions, we evaluated the quality and comprehensiveness of the collected resources. We discarded the papers that did not provide sufficient details to be utilized in our investigation. Regarding this matter, we devised a checklist that contained the following queries:

Q-1: Are there datasets used related to diabetes prediction?

Q-2: Are Machine Learning techniques for diabetes prediction clearly defined?

Q-3: Are there training and validation strategies used for the models?

Q-4: Are there any metrics used to evaluate models for diabetes prediction?

There are three possible responses to each question: **"Yes," "Partially,"** and **"No."** To evaluate the quality and comprehensiveness of each source more accurately, we assigned a numerical value to each label. For example, the label **"Yes"** was assigned the value **"1,"** **"Partially"** was assigned the value **"0.5,"** and **"No"** was assigned the value **"0"** The overall quality score was determined by adding the scores of the responses to the two questions, and the articles with a quality score of at least one was accepted for publication.

Therefore, all the 53 papers that went through previous rounds of evaluation also passed through the quality assessment test. There was no paper omitted in this stage since all the papers were found to have reached the minimum quality score for the next stage of analysis. This phase allowed confirming that the last set of studies was both complete and of quality for the systematic review, thus, guaranteeing that the included studies would advance meaningful insights and reliable information to the research objectives.

### 3.6. Data Extraction

As a part of the research into ML models for diabetes predictions, researchers carefully choose the datasets suitable for training the models, emphasizing their relevance and representativeness [38]. The classifiers usually selected to deal with medical data interpretation issues are developed to cope with the specific difficulties of medical data classification. The training process of ML models often consists of a detailed scheme, which can include cross-validation to ensure that the results are accurate and not overfitted. Validations are performed with test sets of specific data to estimate the generalization of model. Essential metrics such as accuracy, sensitivity, specificity, and AUC measure the performance of model [54]. The choice of independent variables used in training is of paramount importance and could be demographic, biochemical, or clinical factors related to the risk of diabetes. Nevertheless, the research is confronted with restrictions, e.g., biased datasets, the variability of data quality, and the generalization of the results for other communities. These constraints highlight the need for continuous research and overall improvement of ML models in the medical domain. Once we had determined the specific group of sources to be considered, we retrieved the information to answer our research questions. The first thing we did was to specify the data extraction that is displayed in Table 2.

**Table 2**
Data Extraction Form

| Dimension | Attribute Description |
|---|---|
| Datasets | Which datasets were chosen to train the models of ML? |
| Classification Types | Which classification algorithms were selected for diabetes prediction? |
| Training Strategy | What strategy was followed by the ML models for training? |
| Independent Variables | Which independent variables were selected during the model training? |
| Validation | What type of validations were performed for the evaluation of the ML model? |
| Evaluation Metrics | Which metrics were considered for the evaluation of the ML model? |

We also sought to extract data on the datasets exploited and the training and validations used to develop the technique. These facts could help enhance the picture of the chosen features of the papers. In addition to the fundamental information on the diabetes prediction topics discussed in the article or the prediction techniques of ML, we also sought to extract data from the datasets. In addition, the data extraction sheet allowed us to fill in the "*Limitation(s)*" column with the limits identified for the reviewed research methodologies.

## 4. Results and Discussion

Before embarking on the analysis of findings from our systematic literature review, it is prudent to anticipate a systematic approach that allows for accurate interpretation and synthesis of the collected data. This preparatory step requires carefully classifying all collected articles based on criteria such as study design, methodological approaches, ML applications, and effect measures. It is an efficient method of organizing information, which not only helps in the analysis process but also increases the accuracy and reliability of the results. Secondly, we will explain the specific methods applied in the qualitative and quantitative analysis, enabling the meaningful comparison of results across multiple studies. This will help us to provide a comprehensive analysis of the issues discussed in the review and address the research questions and objectives established in the first stages of the review. Figure 2 shows a fluctuating interest in diabetes prediction research from 2014 to 2023. From 2014, there was a surge in publications, peaking at 8 in 2017. The highest number was in 2021, likely due to technological advancements and COVID-19 pandemic situations. However, the decline in 2022 and 2023 suggests a need for increase in the research field.

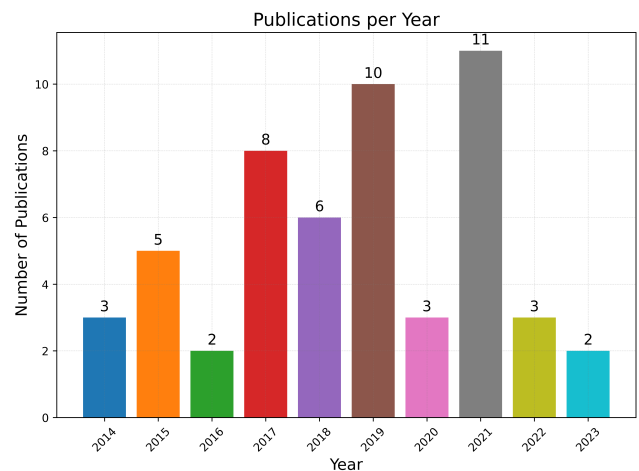

**Figure 2**: Distribution of Publication by Year

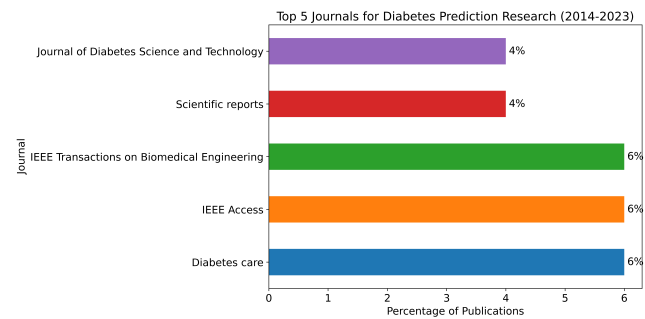

**Figure 3**: Top 5% Journals for Diabetes Prediction Research (2014-23)

Journals play a vital role in disseminating the research done by people. In the context of diabetes prediction research from 2014 to 2023, the top five journals contributing significantly to diabetes prediction research were Diabetes Care, IEEE Access, and IEEE Transactions on Biomedical Engineering. Scientific Reports and the Journal of Diabetes Science and Technology followed closely as shown in Figure 3, accounting for 4% and 4%, respectively. This interdisciplinary approach highlights the growing use of advanced computational methods and data analytics in diabetes prediction.

From our study it is clear that the research on diabetes prediction focuses on critical components such as diabetes, model, data, machine and learning as we can see in Figure 4. The highest frequency of these keywords highlights the importance of data-driven models and machine-learning techniques in enhancing prediction. The research emphasizes the need for accurate predictions to mitigate risks associated with diabetes, utilizing clinical insights and algorithmic advancements.

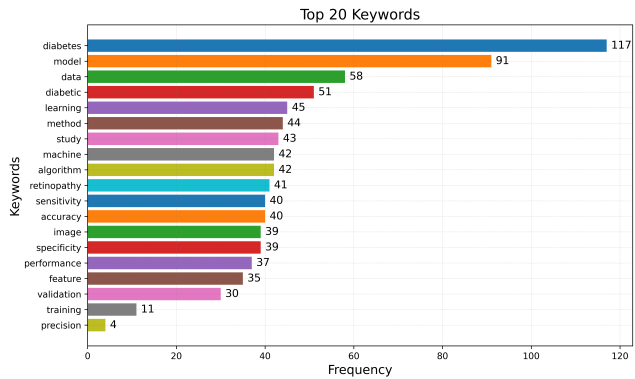

Figure 4: Top 20 keywords used in the studies

### 4.1. RQ$_1$: On the Datasets and Their Characteristics

When it comes to diabetes prediction research, the selection of datasets is rather crucial [37]. Data or datasets are the key input to the development of any predictive model and the quality of the data defines the efficacy of the resulting model. Advanced datasets include people of different demography, diseases, and geography; that provide extensive information for diabetes control and prognosis. From the different datasets, one can see that the way diabetes is researched and combated is quite varied in terms of methods and strategies. In response to the **RQ$_1$**, the current studies revealed that the employed data embraces different populations, and geographic areas, which offered a comprehensive view of diabetes research from different perspectives. Both datasets require different features, which represent the richness and complexity of diabetes management and prediction. Different distribution of datasets used in the studies is shown in Figure 5.

**Longitudinal Studies:** The Japanese study of Aizawa Hospital from Matsumoto investigated 2,105 cases of adults with prediabetes follow-up data with an average observation period of 4. 7 years [79]. It also shows that the monitoring process is a vital aspect for one be able to notice the transition from prediabetes to diabetes. This way, the researchers would be able to determine potential early indicators and contributing factors to the development of diabetes, which could be of significant help in the development of primary prevention and early detection programs. The studies carried out at the Almazov National Medical Research Center cover patients with endocrine disorders, including gestational diabetes mellitus (GDM) [6, 78]. This study is important since it deals with the specifics of diabetes and the challenges that it has on pregnancy in particular. Thus, analyzing the clinical information of this population, the study intends to advance the management approaches and performances in this field for both the mothers and their babies, stressing the importance of focusing on endocrine disorders and developing more specific care for women.

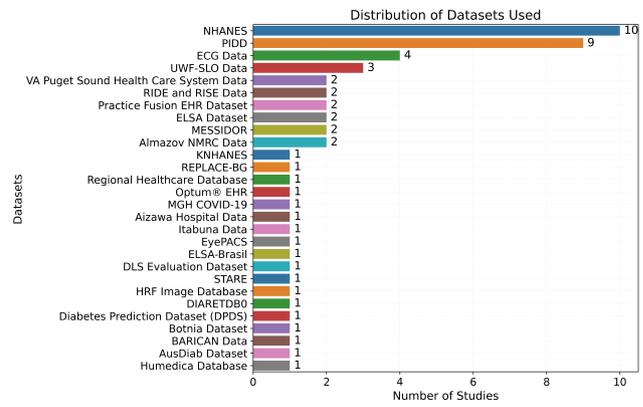

Figure 5: Distribution of Datasets Used in the Studies

**Large Population-Based Studies:** The Australian Diabetes, Obesity, and Lifestyle AusDiab study is a large, cross-sectional study that is ideal for assessing the performance of risk models for diabetes complications using real data of patients [66]. Researchers themselves should have this dataset based on which they could work out and check good models for prognosis of the further evolution of diabetes in patients. The extensive follow-up and the frequent collection of data offer a vast source of information on the relation of such factors as obesity and diabetes. The eight-year follow-up of the BARICAN cohort deals with impact of RYGB on diabetes in 175 patients with T2D. This study gathered data before surgery and at various intervals up to the 5-year follow-up post-surgery to analyze the impact of surgical procedures on diabetes remission and other metabolic outcomes. This research also presents the probability of the utilization of bariatric surgery as a solution for type 2 diabetes.

**Specific Health Condition Datasets:** The Diabetes Prediction Data Set (DPDS) contains data of 520 patients and is used for classification problem, namely, to diagnose a patient as a diabetic or a non-diabetic based on the age, gender, BMI, and lab test results of patient [7]. This dataset is very vital in the development of machine learning models for diagnosis of diabetes; this explains the usefulness of data mining in diagnosis of some diseases. The richness of attributes in the DPDS allows for such a detailed picture of the determinants of diabetes to be painted.

**Imaging-Based Studies:** The Diabetes Prediction Data Set DPDS is a database consists of 520 records of patients that aims to classify them as Diabetic or Non-Diabetic based on their age, gender, BMI and laboratory tests results [7]. This dataset is useful in creating machine learning algorithms of diabetes and shows how data mining is useful in the diagnosis of illnesses. Due to the rich list of features offered by the DPDS, it is possible to get comprehensive information about determinants of diabetes.

**Cardiovascular Metrics:** The baseline of 20 normal and 20 diabetic subjects having 71 records, and 1000 samples each would be used for discriminating normal and diabetic HRV signals [71, 35, 30]. This binary classification problem demonstrates that cardiovascular measurements can detect diabetes, contributing to non-contact diabetes screening and managing.

**Elderly Population Studies:** ELSA follow-up data project of aging office workers can predict type 2 diabetes. It is necessary to screen high-risk elderly persons for diabetes in order to detect them and provide suitable interventions. In the cross-sectional study of ELSA-Brasil with a sample of 15,105 adults, diabetes and cardiovascular health concern were identified [58]. This study establishes health risks such as silent diabetes and advises the respondents to undergo regular health check-ups through structured interviews and laboratory analysis.

**Comprehensive Imaging Datasets:** EyePACS database has 22,075 color fundus images from diabetic patients of different stages of DR [36]. Ethnic variation models and camera models enhance the diagnosis accuracy. Thus, the current work on photo-detection of retinopathy stage develops screening methods for various patient bases. Within the year 2007–2012, 24,331 records of patients from Humedica database were used to identify transitions between normoglycemia, T2D, and prediabetes using the algorithm by Anderson et al. This large database enables the identification of the subjects at risk of developing diabetes or who need closer monitoring.

**Mobile Diagnostic Tools:** Findings from the cross-sectional study involving 824 type 2 diabetic patients from the Itabuna Diabetes Campaign show that there is an inequality regarding DR phases [52]. Equipment used in this study for fundus imaging includes portable retinal cameras, and this proves that other diagnostic equipment can be portable. This simply implies that through the use of modern technology and advancement in the provision of healthcare, patients diagnosed with diabetes can easily access treatment even in remote areas.

**National Health Surveys:** Screening models were estimated from the KNHANES data of 2010 and 2011 for prediabetes prevalence. This is the advantage of the study since it applied large, nationally representative samples to reduce bias and enhance generalizability, highlighting the significance of public health surveys in diabetes research. The NHANES database identifies diabetes in 500 persons by using the demographic, laboratory, and medical history data [75, 45, 60, 64, 70, 47, 35].

**Specialized Databases:** As stated by [42] and [20], the MESSIDOR database categorises normal and DR based on 1200 coloured retinal fundus images The diagnosis technique is evaluated using this data base with the number of photographs and proximity to the three ophthalmological departments. In this paper, we analyzed the effect of COVID-19 on the comorbidities of patients in the hospital using the MGH COVID-19 Data Registry. In goals of ICU admission, mechanical ventilation, and 14-day hospital mortality, chronic factors influence acute ones. A large-scale study using a deidentified Optum® EHR dataset of 95,823,300 patients may be used to predict type 1 diabetic DKA [74]. From a large data set, big data analytics is able to distinguish DKA cases and controls, thus showcasing its diabetes management capability.

**Widely Used Research Databases:** There are numerous studies that have used the PIDD dataset, which consists of 768 female patients with various characteristics, to predict diabetes [63, 64, 46, 3, 72, 56, 80, 44]. Nevertheless, this dataset remains beneficial in diabetes research because it is possible to diagnose diabetes based on pregnancies, glucose levels, and BMI.

The diabetes of next year onset is forecasted using deidentified information from 9,948 patients from the Practice Fusion EHR dataset [55, 19]. The prediction of diabetes development based on the 2009–2011 data to be the situation in 2012 is an example of how EHR can contribute to the longitudinal health monitoring and early detection. One thousand diabetic and non-diabetic patients from a regional healthcare database can be applied for binary classification tasks to predict diabetes [65]. This study is localized and reveals the prevalence rate and risk factors of diabetes within a certain community, thus highlighting the significance of regional investigations on diabetes trends.

**Advanced Monitoring Technologies:** Categorising 226 data streams of CGM data for patients with type 1 diabetes using the REPLACE-BG dataset [39]. The focus of this study on cgm data is an indication of the modern diabetes care technology and real-time monitoring. The images of Kaggle Diabetic Retinopathy challenge and the data from RIDE/RISE clinical trials are used for two-way classification tasks with CST and CFT [27, 8]. Such studies also focus on the significance of clinical trial data in the determination as well as validation of the diabetic retinopathy diagnostic thresholds.

**Global and Longitudinal Studies:** Use of UWF-SLO images from 9,392 patient's worldwide to assess diabetic retinopathy grading. The classification of diabetic retinopathy is evident with benchmark experience of over 10 years and 311604 retinal examinations. The study also concludes that to increase the diagnostic tool accuracy, tight grading should be employed when using regraded arbitrated datasets for validation [72, 53, 49, 35].

> **Summary of RQ$_1$**
>
> Several data enhance diabetes prediction but they have their drawbacks such as quality, data acquisition, demography, and privacy. Substantial numbers of datasets fail to have follow-up data and also do not use uniform parameters for diabetes. It is necessary to improve the mechanism of creating prediction models through the use of integrated data and more refined algorithms.

## 4.2. RQ$_2$: On the Configurations of ML Algorithms in Diabetes Prediction

The application of ML techniques has enhanced the prediction of diabetes and the reliability of predictions. Due to the exposure to different datasets, ML systems can handle large amounts of data, discover complex and sophisticated patterns, and enhance the accuracy of the outcomes. These algorithms entail independent variables and training strategies. Training processes allow for tweaking and verifying the ML models, while other factors that are outside the training process, such as the demographic data of the patients and the medical statistics, provide input data. Diabetes can also be predicted with the help of machine learning algorithms to analyze big data sets and patterns that are unnoticed by statistical means. The common ML algorithms used for prediction/classification purposes are Decision Tree Classifier, Naive Bayes, Linear Regression, Logistic Regression, K-Nearest Neighbor, CNN, SVM, and XGBoost are suitable for organizing various sorts of data and providing accurate predictions or classifications [48]. These algorithms employ complex inputs of data such as medical images, or physiological readings to enhance diabetes diagnosis and care.

Since the diabetes prediction models require independent variables as the inputs for the algorithm training, some of these variables include age, gender, blood glucose levels and the retinal images. The accuracy of predicting diabetes is influenced by independent variables [14]. Choosing appropriate and full variables is useful for the creation of predictions by the ML algorithms. Training methods are essential when it comes to the enhancement of the ML model. Cross-validation, data augmentation, hyper parameter tuning and selection of features improve models and avoid overfitting. Cross-validation is used to validate the model on different data subsets to make it more reliable. The augmentation of data enhances the generalization of the model since the training data contains variation in the new data set. Feature selection helps in removing the irrelevant features hence reducing the noise and improving the models.

### 4.2.1. The Role of Independent Variables and Training Techniques in Diabetic Prediction Using ML Algorithms

Of the studies, some aimed at the diagnosis of diseases such as diabetic retinopathy, possibly glaucoma, and AMD, where the retinal image was the main independent factor. The type of ML algorithm that has been employed mainly for these tasks is the CNN which is used for image analysis and classification. The training plan relies on giving numerous retinal images to the deep learning system. For instance, the research based on the Singapore National Diabetic Retinopathy Screening Program as well as other multi-ethnic population-based studies further optimized their classifiers with large databases of retinal images to achieve high levels of classification accuracy for such diseases [73, 79, 22, 66].

In REPLACE-BG studies, the principal independent variables were the various glycemic indices such as mean blood glucose level and time in range. Classification was often performed by SVM because this was efficient for small sample sizes and avoiding the problem of over-learning. The training strategy included data partitioning into the training and test sets and the features selection by recursive feature elimination. This method allowed for the appropriate utilization of features that enhanced the model performance [39].

A wide range of demographic and health-related predictors were included in the present studies, and data came from the NHANES dataset; most statistical tests applied were logistic regression. In the classification, emphasis was made on biomarkers that would help distinguish between prediabetes and DM. Training strategies employed comprised five-fold cross validation so as to prevent overtraining of the models; large demographic and health data could then be used to accurately predict diabetes risk [75, 36].

The dataset concerning the Itabuna Diabetes Campaign involved determining the DR severity from fundus images using deep CNNs like PhelcomNet. The process of training also involved some augmentation where the images were rotated and brightness was changed in order to get the best results. These works were expected to enhance the CNN diagnostic abilities on a large dataset of fundus images [12, 68].

The features that were detected when working with studies that employed the Optum® EHR Dataset; XGBoost was commonly utilized because of its capacity to accommodate big data. Training activities included selecting the features and hyperparameter tuning with the help of five fold cross validation. This approach helped to deal with the large EHR data that resulted in the prediction of diabetes and related diseases [50, 60, 64, 71, 10, 7].

The EyePACS was mainly consisted of fundus images and CNN was used for diagnosing and categorizing DR. Training practices included data augmentation and cross-validation that allowed the model to recognize the different DR stages across the population [13].

The studies based on data of ELSA-Brasil used random forest algorithms because they are designed to work with high dimensions and also give the probability of variable importance. The training included parameter optimization and selecting the most appropriate variables with the help of the wrapper methods, demographic, and clinical factors to predict diabetes risk [18].

In the Botnia Prospective Study, for prediction of the type 2 diabetes risk by using metabolomics profiles, regularized least squares regression was used. The training strategy used in this study raised the model generalization and predictive accuracy, including multivariate logistic regression and repeated nested cross-validation [49]. Models were validated with the VA Puget Sound Health Care System dataset in terms of sensitivity and specificity to DR detection with FDA (Food and Drug Administration)-approved models. The training strategy was to use models trained on other datasets and testing on this dataset without retraining to prove the transferability and the robustness of those ML algorithms [22, 35].

From the NHANES data of Korea, prediction of prediabetes was done based on fasting plasma glucose level. The machine learning approaches, ANN and SVM, first applied the grid search and then used 10-fold cross-validation to fit the models and recognize the prediabetes within the population correctly [72].

Consequently, based on the visual analysis depicted in Figure 6, we are provided with a rather obvious realization that age variable is used as the independent variable in the vast majority of the studies to a significant extent. Looking at the process of forecasting diabetes, there are some factors that include but are not limited to the body mass index, blood pressure, glucose level, cholesterol level, insulin level, family history of diabetes, physical activity and diet patterns. Moreover, some of the researchers pointed out that, in the training procedure, the researchers must pay attention to and choose the independent variables in order to improve the accuracy of the forecast. Some of the sub-indices that are grouped under the 'other category' depicted in Figure 6 include: environmental factors, social factors, pharmaceutical use and the likes.

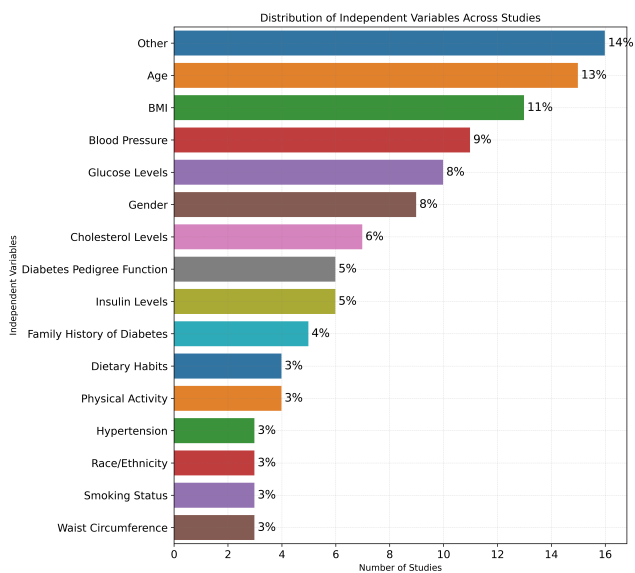

**Figure 6**: Distribution of Independent Variables Across Studies

### 4.2.2. Classification Types and Corresponding ML Algorithms for Diabetes Prediction

CNNs were mainly applied for image analysis in experiments, and the most common type of images used were the retinal ones for diagnosing diabetic retinopathy, suspected glaucoma, and AMD. Training was done using a large number of retinal images and this gave high accuracy models [73, 79, 22, 66].

In analyzing glycemic metrics, Support Vector Machines (SVMs) were used to analyze the small data points with an added advantage of avoiding overfitting. Filtering was the most used technique for adjusting the feature list and extracting the best set of features to be applied in the model [39].

The motivation for applying logistic regression to the NHANES dataset is based on previous application of the method in research on factors such as BMI percentiles and family history of diabetes and hypertension. The process of five-fold cross-validation was a common training methodology to obtain high model stability [75, 36].

Deep CNNs were applied in the Itabuna Diabetes Campaign dataset to classify the DR severity, training strategies were Data augmentation to increase the transferability of model [52, 13].

Specifically, XGBoost was often adopted in research works with the Optum® EHR dataset, which is characterized by high performance and data compatibility. The training methods used were feature selection preprocessing and five fold cross validation [50, 60, 64, 71, 10, 7, 4].

In the studies that employed the ELSA-Brasil data, random forest algorithms were incorporated. These algorithms were selected based on their capacity to work with large number of predictors and give the quantitative evaluation of the importance of variables [18].

In the Botnia Prospective Study, Least Squares Regression analysis was used in the determination of type 2 diabetes risks related to metabolomic profiles. Training methodologies employed were multivariate logistic regression, and repeated nested cross-validation [49, 72, 7].

As for the HRV signals obtained from ECG, CNN-LSTM-SVM Hybrid Models were utilized in the research, since component algorithms complement each other, improving the predictive potential. [63, 68, 70, 23, 55, 47, 42, 35, 28, 18, 8, 69, 56, 61].

The deep learning model Inception-V3 was used, which is ideal for image data, was applied in a study that involved the analysis of OCT measurements of diabetic patients [27].

Demographic, clinical characteristics and lifestyle factors were used and employed with Bayesian networks, as this approach is suitable for representing probabilistic dependencies between variables [58, 6, 3, 72, 53, 43, 41, 30, 20, 80, 32, 57].

Traditional ML and deep learning algorithms are applied in half of the studies on diabetes prediction, thus, proving that they play the equal and significant parts. Logistic regression and decision trees are basic approaches of machine learning, while CNN is suitable for big data. Ensemble

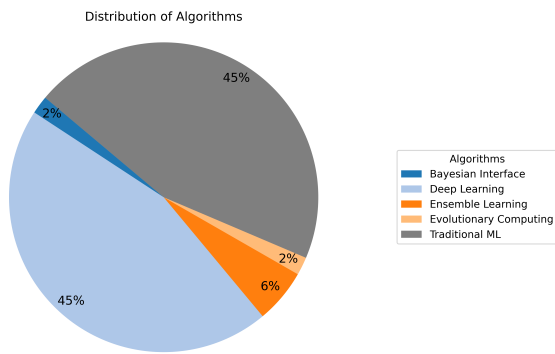

Figure 7: Percentage of ML Algorithms used for Diabetes Prediction

Learning that utilizes more models to enhance performance makes up 6% of approaches. Evolutionary Computing is less than 2% and Bayesian Inference also less than 2% indicating that several methods are used for enhancing the accuracy and complexity of diabetes prediction. Figure 7 shows that the majority of works mainly employ basic machine learning and deep learning algorithms. However, the authors are aware of the existence of many other algorithms in the field including ensemble methods, evolutionary algorithms and Bayesian inference, which indicate the variation of the approaches taken in the diabetes prediction.

> **Summary of RQ$_2$**
>
> Considering the data type and their appropriateness for certain tasks, the algorithms used for diabetes prediction include CNN, SVM, and XGBoost. Cross validation, data augmentation and feature selection helps to increase the convergence of the predictive models, which can also demonstrate the versatility of the different machine learning frameworks in altering the variables, algorithms and training paradigms for diabetes prediction researches.

### 4.3. RQ$_3$: On the Evaluation Techniques and Metrics

The evaluation setups are important especially in diagnostics of diabetes through ML algorithms. It is important to have evaluation sets to make sure the models are correct and practical in real life situations. They provide the robust approach to compare the ML models and thus it is easier to identify the most appropriate techniques and increase the reliability and efficiency of models. The primary finding of this study is that the configurations of ML model evaluations affect the reliability and robustness of the models. Overfitting, check whether the model runs properly on other data, and describe advantages and limitations of the model. Scholars can determine which diabetes prediction models to select based on the available evaluation methods.

As it is seen in the diabetes prediction literature, there are different approaches for measuring the performance. One such method is the cross validation which is helpful in enhancing the assessment of the model by checking for its performance on different partitions of data. This approach makes the model more reliable and accurate when tested on different datasets hence giving more reliable evaluations [39, 75].

Validation tests the model on a data other than training data. It is also important in the validation process to see the generalization of the model on new data. In the Optum® EHR dataset research, the model was externally validated by comparing the prediction with the scores of new images from the screening program as well as ten other datasets with different populations [45]. Bootstrap sampling involves taking at random a number of sample from the dataset, using this sample many times to train the model and obtain an empirical distribution of the performance measures in order to analyze model variability.

It is essential to know the types of assessment criteria because they help quantify the performance of the ML models. They include accuracy, Area Under the Curve (AUC), sensitivity, specificity, precision and F1 score that gives a detailed performance view of the models. These metrics assist in determining the models that not only forecast diabetes with high accuracy but also approach the issue of false positive and false negative cases, which are costly in practice [75, 39].

The methods used in evaluation setup in diabetes prediction research are elaborate to ensure that the performance of the models is tested comprehensively with different approaches and measures.

**Validation Methods:** At high reliability, the data sets are divided into portions and the most common technique used is the k-fold cross validation. This technique divides the dataset into k portions and constructs the model k times, and the validation data set is one of the k portions while the training data set is the remaining portions of the dataset. For instance, the authors of the research conducted on REPLACE-BG dataset applied the 10- fold cross validation to check the efficiency of the SVM model and to avoid obtaining the performance indicators that would not reflect the performance of the model [39]. Similarly, the research that used NHANES data used five-fold cross-validation to verify the overall accuracy of the logistic regression model, however, this method needs more training data [75]. This method is very vital in preventing overfitting, whereby the model gives excellent results on the training data but poor results on the test data, giving a better evaluation of the model.

External validation is performed by using different dataset than the ones used to train the models to infer the generality of the models. It is helpful to evaluate the model in conditions closer to real-world since cross-validation does not show all the aspects of performance. For instance, in the study that relied on the Optum® EHR dataset, external validation was conducted using the comparison of the predictions with the scores assigned to new images from the screening program and 10 other datasets comprising other populations [45]. This type of validation means that the model is applicable to make predictions under different populations and conditions; or in other words in different clinical settings.

Bootstrap sampling is adopted in some research to assess the extent of fluctuations of the performance markers of the model. In this method, a dataset is applied such that it samples randomly and successively with replacement and feeds these samples to the model until the empirical distribution of the performance measure is obtained. Similarly, the identified study that was conducted on the Optum® EHR dataset to identify genetic variants associated with diabetic ketoacidosis (DKA) also used 1000 bootstrap samples to calculate the 95% confidence interval for all the aforementioned performance parameters; thus confirming the authenticity of the statistical measures used in the study [45]. This technique enables the analysts to identify the various levels of variability of the model and the stability of the model in the sampling distribution.

**Evaluation Metrics:** Precision is one of the most broad and the most often used measures in the analysis of investigations to describe the proportion of the actual positives and actual negatives regarding the total number of the investigated cases. For instance, while validating the logistic regression models, the measure used was the accuracy rate which has been obtained in the NHANES-based studies [75]. Other works also considered accuracy as the criterion for the performance of their models, with specific distribution percentages of each [39, 45, 22, 18, 6, 71, 3, 53, 42, 35].

AUC is relevant when comparing binary classifiers as it offers information on the ability of the classifier to distinguish the two classes. It measures the percentage or rate at which it is right in categorizing positive and negative samples. The AUCs obtained were high, which pointed to adequate diagnostic capability of the models for DR using the EyePACS dataset. This metric allows one to compare a true positive rate (sensitivity) with a false positive rate and get a general performance figure [75, 13, 58, 72, 49, 60, 27, 64].

The sensitivity or true positive rate and specificity or true negative rate indicate the capability of the model in the identification of positive and negative cases respectively. Sensitivity measures the true positive rate and specificity is a measure which gives the ability to identify actual negative cases. For instance, the study conducted on data collected in Itabuna Diabetes Campaign indicated that the sensitivity for the screening model was 97%. For detecting the more than mild DR, the model had the sensitivity of 8% and specificity of 61%. 4% in detecting severe cases, suggesting that the proposed model is capable of raising awareness on severe cases of DR while at the same time pointing out the features that require improvement [73, 52, 50, 79, 16, 63, 66, 46, 23, 55, 43, 28].

In the case of dealing with data mining in imbalanced datasets, some of the measures that are considered to be very crucial include precision which is the total number of correct predictions of the positive observation over the total number of positive observations in the data set and F1 score which is a weighted average of both the precision and recall. The evaluation of the SVM model in the study with the REPLACE-BG dataset incorporated these measures, where not only true positives are correctly identified, but where the proper precision and recall of the model is achieved as well. [39, 36, 58, 68, 70, 41, 19, 4, 44, 32, 57]. These are good metrics particularly when it is necessary to avoid the position where the model provides both high false-positive and high false-negative values.

The measures of diagnosis accuracy are Positive Predictive Value (PPV) and Negative Predictive Value (NPV), which reveal the proportion of actual positives and actual negatives out of all the cases predicted to be positive or negative. The study that used the data from VA Puget Sound Health Care System with the purpose of comparing the effectiveness of the screening algorithms developed with the help of AI and with the help of PPV and NPV results in giving more understanding regarding the efficiency of the models in the actual health care centers [22]. These metrics are useful when a model is applied in a clinical context in which false positive and false negative results can have consequences.

Figure 8, shows the various types of measures that are employed for assessing the diabetes prediction models. Looking at the results, accuracy stands out as the most commonly employed measure, which only further emphasizes the significance of the measure in evaluating the models' performance. Sensitivity and specificity are also prominently used, and hence, the equal emphasis on detecting both positive and negative cases. This just goes to show that the use of AUC is important in evaluating the discriminative capability of the models. Precision which calculates the true positive predictions and F-Measure which balances precision and recall are also important but not as popular as the above mentioned. It is worth mentioning that PPV and NPV are among the least used metrics, which implies that, though crucial, they are not the major concern for most investigations. In sum, the figure portrays a detailed approach to model evaluation, although the author leans toward measures that offer a general picture of the model's performance.

> **Summary of RQ$_3$**
>
> This systematic review validates the methods and metrics used in the prediction of diabetes using machine learning across the spectrum. To enhance the model reliability, cross-validation, external validation, and bootstrap are used; whereas, for checking the model effectiveness, performance evaluation metrics such as accuracy, AUC, sensitivity, specificity, precision, and F1 measure are employed.

### 4.4. Discussion

Drawing from the the studies analyzed with the performed SLR, this paper identifies the important elements that define the process of developing accurate predictive models for diabetes. The discussion covers the choice of datasets and their quality, the chosen machine learning algorithms and training paradigms, and the evaluation scenarios and measures used in the assessment of the performance of the model. In this review, the strengths and limitations of the different approaches are discussed based on many publications, paying attention to the issues of diversity and standardization of datasets, feature selection and preprocessing, and evaluation methods. Thus, the discussion of this study seeks to provide an understanding of the status of diabetes prediction research and define the directions for its future enhancement to create more precise, accurate, and universally applicable predictive models.

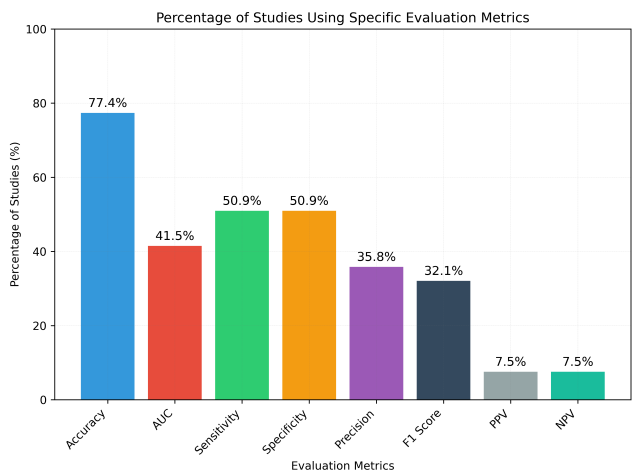

Figure 8: Percentage of Studies Using Specific Evaluation Metrics

**Dataset Utilization and Insights:** While addressing $RQ_1$, it was necessary to initially analyze different datasets that are commonly employed in the studies aiming at diabetes prediction. As indicated, the reviewed studies used different datasets from a longitudinal study like Aizawa Hospital to large population-based studies like AusDiab. These datasets are indeed valuable sources of information on diabetes and its prediction, as the nature and variety of this disease indicate. Nevertheless, problems like variable quality of data and irregular approach to its collection point to the issues of weak data standardization. Combining datasets that were obtained from different regions or populations can help to avoid creating models that do not work in different populations and under different clinical conditions.

**Dataset Selection and Quality:** Our SLR shows the great variety of datasets used in diabetes prediction studies proving the value of diversification and dataset samples. The characteristics and challenges of each dataset are different. For example, NHANES has a large and racially/ethnically diverse sample with extensive demographic, clinical, and lifestyle data. However, problems like class imbalance where one class has many samples and the other has few become a problem for the model performance and its ability to generalize.

**Data Quality and Consistency:** It is more important for the quality of the datasets that go into the building of these models and for their consistency. Probable sources of bias include differences in data collection techniques as well as dissimilarities in the definition of diabetes used in the different studies. Some of these problems can be obviated where the data collection protocols are standardized. For instance, the cross-sectional study of the patient database of Aizawa Hospital, Japan, and the large population base cross-sectional AusDiab study shows that the methods of data collection may differ. This limitation is even compounded by the fact that many of the studies available do not provide follow-up data, thus limiting the possibilities of assessing the long-term value of predictive models. In addition, the issues related to heterogeneity in the diagnosis and assessment of diabetes-related variables including fasting glucose levels, HbA1c thresholds, and diagnostic criteria also make it challenging to synthesize evidence from different studies.

**Standardization and Integration:** It is crucial to note the attempts to establish international guidelines for the collection of data and reports concerning diabetes. To increase the reliability of prediction models, it is necessary to unify the methods of data collection and combine different datasets. This approach can assist in addressing the current drawbacks like inconsistency in demographic and clinical variables that can influence the results of the model. Combining datasets from different regions and population groups may offer a broader understanding of diabetes and thus enable the creation of models that are viable in different populations and clinical situations. Additionally, the combination of EHRs, genetic information, and CGM data with clinical and demographic data can improve the current accuracy and comprehensiveness of the model.

**Machine Learning Algorithms and Training:** The $RQ_2$ was on the approach used in the machine learning algorithms in predicting diabetes as well as the training approaches with independent variables. The studies showed how these algorithms such as CNNs, SVMs, and XGBoost perform with different data types and with different types of predictions. This is to show that independent variables and training strategies like cross-validation and data augmentation were critical to improving the performance of model. Whereas the choice of features and training methods influenced the effectiveness and transferability. The use of these ML algorithms in different studies proves the efficiency of these algorithms in giving accurate predictions in different clinical areas.

**Algorithm Selection:** The SLR reveals several ML algorithms utilized in diabetes prediction; they are CNNs, SVMs, Logistic Regression, and XGBoost. Every of them has its advantages and is used for different data and prediction problems. For instance, CNNs are efficient in analyzing the retinal images for DR, whereas SVM is employed for analyzing the glycemic indices and demographics data. Despite the presence of more complex models, logistic regression continues to be used due to its simplicity and ease of interpreting the results while analyzing structured clinical and demographic data. XGBoost is preferred due to the demonstrated superiority with tabular datasets, and the flexibility in handling missing values and feature interactions.

Table 3
Key Findings from RQ$_2$: On the Trainings Strategies, Independent Variables and ML Algorithms

| Studies | Independent Variables | Training Strategies | ML Algorithms |
|---|---|---|---|
| [73, 79, 22, 66] | Retinal images | Cross-validation | CNN |
| [39] | Glycemic Indices | Recursive feature elimination, training/testing split | SVM |
| [75, 36] | Demographic and health-related variables | Five-fold Cross-validation | Logistic Regression |
| [12, 68] | Fundus Images | Data Augmentation | CNN (PheIcomNet) |
| [50, 60, 64, 71, 10, 7, 4, 65, 44] | EHR data | Feature selection, five-fold cross-validation | XGBoost |
| [13] | Fundus images | Data augmentation, cross-validation | CNN |
| [18] | Demographic and clinical features | Parameter tuning, wrapper approaches | Random Forest |
| [49] | Metabolomics profiles | Multivariate logistic regression, cross-validation | Regularized Least Squares Regression |
| [22, 35] | Retinal images | Testing on new dataset without retraining | Various FDA-approved models |
| [72] | Glucose Levels | Grid search, 10-fold cross-validation | ANN, SVM |
| [63, 68, 70, 23, 55, 47, 42, 35, 28, 19, 8, 69, 56, 61] | HRV signals from ECG | - | Hybrid Models(CNN-LSTM-SVM) |
| [27] | OCT measurements | - | Inception-V3 |
| [58, 6, 3, 72, 53, 43, 41, 30, 20, 80, 32, 57] | Demographic, clinical, lifestyle factors | - | Bayesian Network |

**Feature Selection and Data Preprocessing:** The selection of independent variables is appropriate when developing these models. Inputs that are required are demographic data, clinical measurement, and medical images of patients. Some of the critical techniques in the development of any model include feature selection and data preprocessing. Some of these are the choice of features for CNNs for retinal image analysis and the training techniques for SVMs for glycemic indices. Feature selection techniques like the Recursive Feature Elimination and Wrapper methods guarantee that only the variables that are most beneficial to the model are used, thus decreasing the noise level. Other preparations that may help to enhance work on the project include data scaling, handling of missing values, and converting categorical data to numerical data.

**Training Strategies:** Training strategies like cross-validation and data augmentation and feature selection are important for increasing the model reliability. Complexity and accuracy, as well as the independence of the data, are checked with methods like k-fold cross-validation to avoid overfitting. Data augmentation especially in image-based investigations enhances the performance of model because of the variations in the data set. For instance, the variation in the retinal images by rotation and changing the brightness of the images can improve the generalization of the model. Pre-processing needs involve feature selection techniques, which help identify the best variables to use in the model eliminating the noisy ones. Bagging and boosting are other techniques that are also used to enhance on the performance of the models by using multiple models and coming up with one final result that is more accurate as compared to the individual models.

**Applications and Use Cases:** The same ML algorithms have been applied in many research studies implying their efficiency and usefulness. For example, CNNs have been applied in identification of Diabetic retinopathy from the retinal images, SVM in analysis of ECG data for Diabetic and Non-Diabetic Heart Rate Variability and XGBoost to large scale EHR data for diabetes onset and complication prediction. The above applications prove that with the execution of the ML algorithms, predictions made are accurate and reliable in different clinical practices. Further, incorporation of the ML models with the clinical decision support systems (CDSS) may help the clinicians to make better and timely decisions which may help in improving the quality of life of a patient and decrease the impacts of complications due to diabetes.

Table 4
Key Findings from RQ$_3$: Evaluation Techniques and Metrics

| Studies | Evaluation Setups | Evaluation Metrics |
|---|---|---|
| [39, 75, 45] | 10 fold Cross Validation, External Validation | Accuracy, Precision, Recall, F1 Score |
| [22] | External validation | PPV, NPV |
| [18, 6, 71, 3, 53, 42, 35, 10, 69, 65] | 5-fold cross validation | Accuracy |
| [13, 58, 72, 49, 60, 27, 64, 70, 7, 80] | Cross Validation | AUC |
| [73, 52, 50, 79, 16, 63, 66, 46, 23, 55, 43, 28] | Cross Validation | Sensitivity, Specificity |
| [36, 68, 41, 19, 4, 44, 32, 57] | Cross-validation | Precision, Recall, F1 Score |
| [22] | External Validation | PPV, NPV |

**Evaluation Setups and Metrics:** As derived from **RQ$_3$**, which aimed at identifying the evaluation setups employed in the assessment of the machine learning models for diabetes prediction, including the kinds of validation employed and the measures used to measure the performance of the models. The studies used other techniques in order to assess the validity and portability of model where k-fold cross validation, external validation, and bootstrap sampling were used. Measures like accuracy, AUC, sensitivity, specificity, precision, and F1 score offered upto satisfactory measures of model performance and thus, underlined their value in clinical use. These evaluation setups and metrics ensure that models are not only valid in providing a range of clinical applications but also valid in terms of delivering rich and comprehensive information concerning model performances and thus helping to distinguish the most suitable predictive models. Key findings related to RQ$_3$ are presented in Table 4.

**Deciding Evaluation Methods:** The evaluation setups that are incorporated in the diabetes prediction research are aimed at achieving the ML model validity and reliability. Cross validation, External validation and Bootstrap sampling are some of the techniques that offer strong guidelines on the performance of the model. They assist in detecting overfitting, guarantee the generalization of models, and provide information about the advantages and limitations. For example, k-fold cross validation where the set of collected data is split into k sets where one is used for validation data while the other k-1 are used as training data offers a more comprehensive result of the model.

**Selecting Key Metrics:** To assess the performance of the model, the evaluation measures of accuracy, Area Under the Curve (AUC), sensitivity, specificity, precision, and F1 score are essential. These metrics give detailed information on how well a model identifies diabetes, reduces false positive and false negative rates, and works in practice. For instance, accuracy quantifies the number of true results (both true positives, and true negatives) per total analyzed cases, whereas AUC gives information on the ability of model to classify classes. Sensitivity and specificity: These are critical metrics when it comes to analyzing the capacity of model to diagnose positive and negative cases. Accuracy is not recommended for imbalanced datasets since it tends to favor the majority class, while Precision and F1 score which take into account both precision and recall are recommended for use with imbalanced datasets.

**Ensuring Reliability:** The validity of the models can be confirmed through the use of other datasets other than the ones used in training as a way of testing if the models developed will work well on new data. For instance, the study that was conducted using the Optum® EHR dataset did the external validation through the comparison of model predictions to the scores of professional graders of new images from the screening programs, as well as other diverse population groups. Bootstrap sampling, which trains the model multiple times with random samples drawn with replacement from the dataset, gives an empirical measure of the variability of the performance measures. These evaluation setups help ensure that the models that are developed are more reliable, valid, and transportable to a broad spectrum of clinical practice. Also, the interpretability and explainability of the models are essential, especially in the clinical environment, where the ability to understand the decision-making process of the system will increase acceptance by practitioners.

The discussion also draws focus to the selection of datasets, the choice of machine learning algorithms, and the evaluation frameworks in the creation of reliable diabetes prediction models. Hence despite the progress that has been made in the field, solving problems about data quality, data consistency, and data privacy is crucial for future development. Here, interdisciplinary cooperation and compliance with the standardized procedure of data collection and the use of sophisticated algorithms will allow for the potential of machine learning to be realized and contribute to the efficient treatment of diabetes. The use of multiple sources of data, appropriate selection of features, and better training and validation paradigms will improve the robustness and transferability of the predictive models and thus contribute to the betterment of lives as well as the field of diabetes prediction. Nevertheless, *what are the main limitations that have been found in this SLR, and how can future studies deal with these issues to enhance diabetes prediction models?* This question will be discussed in the next section to identify the current limitations of diabetes prediction and future directions for improvement of the models.

## 5. Research Limitations and Implications

Diabetes prediction using Artificial intelligence, specifically Machine learning (ML) has presented a way of early diagnose and effective control of the disease. Our study found some issues such as data quality, feature selection, model complexity, and the ethical implications make it difficult to achieve in the healthcare domain. Solving these problems is vital to build stable and accurate ML models to be incorporated into the clinical setting. In this section, we discuss the limitations and their consequences for researchers and stakeholders and outline the steps to enhance the field and enhance diabetes prognosis and management.

**Data Quality and Availability:** Prediction of diabetes cases involves the use of quality and accessible data. More popular datasets such as PIDD or NHANES may not be viable or have inherent bias and may not work for other population types. This complicates the training and validation of models because of differences in data gathering methodologies and the presence of gaps in data. Furthermore, the data used may not include all the patient diversity, particularly those belonging to the different minorities, therefore the developed models may not generalize well across the patient population.

☞ *Better quality and variety of data, consistent methods of data gathering, and balancing classes in the models are important for improving the machine learning models' dependability and credibility. This is why healthcare organizations and policymakers should ensure the development of large databases with ethnic, demographic and geographical characteristics of minorities; this would increase the relevance of the models used.*

**Feature Engineering and Selection:** Diabetes prediction models generally involve feature engineering and selection since they use demographic and clinical aspects such as age, BMI, blood glucose level, and others. But the disease is complex and long-term models that do not contain genetic and lifestyle factors are very simplified. Feature selection is always a problem; usually, researchers choose features that are either redundant or irrelevant, which is bad for the model.

☞ *Researchers should use genetic and other related factors as well as lifestyle and behavior data in their research and should use dimensionality reduction and feature importance analysis. The clinicians are the main target that has to be involved in the process of model creation and refinement to achieve accuracy and validity. Therefore, stakeholders should promote programs that increase the number of data assets and select better features for predictions.*

**Model Complexity and Interpretability:** The complexity of the models also poses a problem in the prediction of diabetes. Large neural networks are precise, however, they are overparameterized and overfitting for training data. These models are often 'black boxes', which makes it hard to explain the cause of a decision to a patient or another health care worker. This lack of transparency proves to be a drawback in adoption and patient management.

☞ *Models of moderate complexity should be easier to interpret and that explainability should then be applied as an intervention. If efficient models cannot be used, then more complex models should be used in the predictions. Clinicians should believe in usability models to enhance the implementation of clinical practice. Training and other resource supplies to healthcare personnel need to be encouraged.*

**Population Diversity:** It is often discovered that models developed from specific datasets are not efficient for more extensive populations because of the dissimilarities in genetic constitution, conduct and setting, which can be significantly different from the dataset, especially when such datasets comprise age and gender bias. This can result in the creation of biased models which keeps on emphasizing on the health inequalities.

☞ *The training datasets should include samples from different demographic, clinical, genetic, and behavioral backgrounds to include the affected people in the models. Such updates are vital for responding to today's health issues, increasing the variety of available information, and advancing healthcare.*

**Technological and Infrastructural Limitations:** Complex machine learning algorithms in healthcare can be computationally expensive and present issues like integration, privacy, and security and insufficient infrastructures can cause inequity in the distribution of access and care.

☞ *Researchers have to design a model for health care provision, irrespective of the available resources. There must be efforts put into computing assets by the key players and standardization of the same must be made. Governments should allocate resources for technology and training distributions to reduce the disparity of healthcare quality.*

**Lack of Standardization in Training and Validation Approaches:** The choice of proper methods for the training and testing of the diabetes prediction models is not entirely straightforward because of the lack of clear recommendations; indeed, the efficiency of the models and the ease of comparing conclusions from various studies and approaches may vary.

☞ *Data-augmentation and cross-validation should be employed by researchers to boost the model performances. There should be standard procedures on training and validation that should be set by all the stakeholders in order* to minimize the variations between the different studies and applications. It is suggested that principles and best practices for ML models can increase their dependence and make outcomes more similar and comparable; in this case, institutions should collaborate for better research in this regard.

**Inconsistencies in Evaluation Setups and Measures:** The case shows that the standards and settings of ML model assessment and selection are not the same, while quantitative indicators such as accuracy, AUC, sensitivity, and specificity are not applicable in real life, which often does not allow distinguishing between patients with early phase diabetes and non-diabetic individuals, which is important for correct actions.

*The used evaluation metrics should include accuracy, AUC, sensitivity, specificity, and clinically oriented performance. The stakeholders should ensure that standard evaluation tools are used to serve the clients and meet the market standards. Criterion for improvement of the predictive models based on statistical results and clinical relevance will be developed. This will improve the credibility of the ML models since openness will be promoted.*

**Ethical and Legal Considerations:** There are ethical and legal issues that come with ML models in diabetes prediction such as bias in the training data and measures such as Health Insurance Portability and Accountability Act (HIPAA) and General Data Protection Regulation (GDPR) that limit data access. These problems can aggravate health disparity and restrict data protection and security as well as impede data sharing and model building.

*The models that have been developed require high reliability, and the ethical norms such as HIPAA and GDPR contribute to it. This guarantees appropriate use of data making statistical models more precise and helps to eliminate discrimination and unfair practices in the sphere which in its turn will benefit society and provide equal treatment for patients.*

**Lack of Collaboration on Diabetes Prediction Efforts:** Collaboration in the diabetes prediction hamper the data accessibility for constructing a model, its testing, moral analysis, and its practical use in real-world settings, which slows down research, development, and patient outcomes and benefits, while there is a requirement for diverse and non-bias data.

*Clinicians and stakeholders should engage researchers in creating communication channels for new models in order to improve performance and fit the demands of the healthcare sectors. Involvement of stakeholders helps in producing clinically meaningful, technologically realistic, and ethically sound models that enhance patients' experience.*

## 6. Threats to validity

Like any other systematic literature review (SLR), the present study has some limitations that could have potentially affected the validity of the observed results. This section presents these limitations as well as the measures that were taken to manage them.

**Literature Selection:** One of the crucial issues in conducting a systematic literature review is how to find sufficient papers to provide a general understanding of the state of the art of a given research area. In this regard, the present study formulated a comprehensive search question with no temporal restrictions to acquire as many papers as possible concerning the application of machine learning for diabetes prediction. Although this approach was time-consuming, it was used to achieve exhaustiveness. It is important to notice that synonyms and alternative spellings of the terms commonly used in the literature for defining the search query were identified. Furthermore, we looked for these search terms among the systematic literature reviews on diabetes prediction to see if there are other suitable terms. To enhance the data collection in the research area even more, a backward snowballing session was done on the papers obtained after the exclusion/inclusion criteria were applied. To ensure credibility all the processes to arrive at the choice of the primary studies were cross-checked by at least one of the authors. The implementation of these actions allows us to have confidence in the comprehensiveness of the selection of literature sources. To ensure that all the steps and intermediary results of the analyses reported here can be verified and independently replicated, all of them are presented in the online appendix.

**Literature Analysis and Synthesis:** Following the selection process, the following exclusion criteria were used to remove papers that could not make a positive or would make a minimal, contribution in the summarization of the state of the art about the defined research questions. We did not restrict the list of primary studies to the articles that meet the inclusion criteria only but also performed an extra quality check to confirm their relevance. To make sure that no resource that does not meet the objectives of the paper is included, this manual assessment posed an additional layer to the process.

More broadly, the literature synthesis was performed according to the results of the manual analyses, and these are known to be prone to human factors. In this regard, two observations are necessary. First, the two main authors were involved in the process which reduces subjectivity and possible mistakes. Second, the presence of third author was always constant, and he gave inputs on how to perform the different phases of the systematic literature review whenever necessary.

These combined efforts go a long way toward reducing the threats to validity and provide a thorough and comprehensive review of the current state of affairs in the use of machine learning techniques in the prediction of diabetes.

## 7. Conclusion

This systematic review demonstrates the future progress and the productivity of the ML in the diagnosis and management of diabetes, a major global health concern. This way, while comparing 53 studies, the review offers an overview of the datasets, ML algorithms, training methods, independent variables, and evaluation metrics used in diabetes prediction. Some of these datasets include the Singapore National Diabetic Retinopathy Screening Program, REPLACE-BG, National Health and Nutrition Examination Survey (NHANES), and the Pima Indians Diabetes Database (PIDD), which come with their peculiarities, some of which are class imbalance. The review highlights the positive impact of several of the ML algorithms such as CNN, SVM, Logistic Regression, and XGBoost in diagnosing diabetes results. Other attributes that are often used as independent variables include age, body mass index, blood glucose concentrations, genetic polymorphisms, and lifestyles, which are instrumental when building forecasting models. Also, the review gives insights into methods like cross-validation, data augmentation, and feature selection that improve the flexibility and stability of models. Therefore, it is crucial to use such assessment indicators as accuracy, AUC, sensitivity, and specificity to provide a comprehensive assessment of the model.

In the future, it is necessary to overcome the current weaknesses to enhance the utilization of ML in diabetic prediction. Future studies should pay attention to the quality and variability of data, methods of handling the class imbalance, the interpretability of the model, and the computational complexity. The multi-center studies involving various population groups, and standardizing the metrics for evaluation and validation of the models are the few important steps that need to be taken. If these challenges are addressed, then ML has the potential of enhance the accuracy of diagnosis, the health of the patients, and the effectiveness of the healthcare system, hence lowering the global impact of diabetes. In light of such findings, this review calls for the integration of ethicists and other stakeholders in formulating recommendation policies involving the application of ML-based diabetes prediction models aimed at enhancing the quality of life of people globally through the use of AI technology in the delivery of healthcare services. The findings and recommendations of this review are useful in the current drive towards the use of AI and ML in combating one of the biggest challenges to health in the modern world.

## 8. Acknowledgments

This work has been partially supported by the European Union through the Italian Ministry of University and Research, Project PNRR "*D3-4Health*: Digital Driven Diagnostics, prognostics and therapeutics for sustainable Health care". PNC 0000001. CUP B53C22006090001